\documentclass[a4paper]{article}

\usepackage{graphicx}
\usepackage{pdfpages}
\usepackage{mathtools}
\usepackage{multirow}

\providecommand{\keywords}[1]
{
  \small	
  \textbf{\textit{Keywords---}} #1
}

\usepackage[sorting=none]{biblatex}
\begin{document}

\title{Determination of the stress state beneath arbitrary axisymmetric tangential contacts in Hertz-Mindlin approximation based on the superposition of solutions for parabolic contact}

\author{E. Willert \\
Technische Universit\"at Berlin\\
Stra\ss{}e des 17. Juni 135, 10623 Berlin, Germany \\
e.willert@tu-berlin.de}
\maketitle

\begin{abstract}
As an improvement to the recently proposed procedure for the determination of the stress state beneath axisymmetric tangential contacts in Hertz-Mindlin approximation via an appropriate superposition of solutions for the respective flat-punch problem, the determination via the superposition of solutions for parabolic contact is demonstrated. It has two advantages over the flat-punch formulation: the numerical implementation is slightly easier and more stable, due to the absence of stress singularities for the smooth parabolic profile. As a numerical example, the oscillating tangential contact between a rigid indenter with a profile in the form of a power-law (with exponent 4) and an elastic half-space is considered in detail.
\end{abstract}


\section{Introduction}\label{sec1} 

Mechanical contacts are among the most severely stressed components of structures. Nevertheless, as a solution to a given contact mechanical problem, often only the traction vector in the interface is determined. For elastic contacts that obey the restrictions of the half-space approximation, in theory, the complete stress state beneath the contact can be determined from the surface tractions via a superposition of the fundamental solutions for the Boussinesq \cite{Boussinesq1885} and Cerruti \cite{Cerruti1882} problems. In practise, however, this procedure is numerically quite costly. Alternatively, one can apply the potential theory solution provided by Barber (\cite{Barber2018}, Appendix A). For the general axi\-symmetric Boussinesq problem, the stress state beneath the contact has recently been given in integral form based on the Hankel transform \cite{Yangetal2021}. Exact closed-form solutions are usually restricted to the Hertzian contact problem under normal and tangential loading, see the works by Huber \cite{Huber1904}, Sackfield \& Hills (\cite{SackfieldHills1983}, \cite{HillsSackfield1987}, \cite{SackfieldHills1983b}) and Hamilton \& Goodman (\cite{HamiltonGoodman1966}, \cite{Hamilton1983}).

On the other hand, it has recently been demonstrated (\cite{Willertetal2020}, \cite{Forsbach2020}), how the full stress state below the surface in axisymmetric tangential contact problems can be determined via the appropriate superposition of respective flat-punch solutions. This procedure is valid within the restrictions of the Hertz-Mindlin approximation, which comprises the assumptions of elastic similarity of the contact partners, local Amontons-Coulomb friction with a constant coefficient of fricion and negligence of the lateral surface displacements resulting from the tangential contact tractions. The superposition of flat-punch solutions, however, in this context has two drawbacks: Firstly, the pressure distribution for flat-punch indentation is singular at the contact edge, and, accordingly, as are the superposition integrals for the determination of the stress state in the general axisymmetric case. And secondly, the closed-form full solution for the flat-punch problem is only calculated via the (extremely lengthy) derivatives of the known respective solution for parabolic contact.

Therefore, it would rather be preferable, to formulate the axisymmetric solution directly in terms of an appropriate superposition of the solution for parabolic contact. How this can be done, is demonstrated in the present manuscript. First, the full superposition procedure is laid out. After that, a numerical example of an oscillating tangential contact is studied in detail, and some conclusive remarks finish the manuscript.

\section{Superposition procedure}\label{sec2}

\subsection{Normal contact}

Consider the contact of two linearly elastic bodies with the shear moduli $G_1$ and $G_2$ and the Poisson ratios $\nu_1$ and $\nu_2$. Both shall obey the restrictions of the half-space approximation, i.e., their macroscopic dimensions are much bigger the largest characteristic contact length and the surface gradients in the vicinity of the contact are small. The point of first contact shall be the origin of a cartesian coordinate system $\left\lbrace x,y,z \right\rbrace$, $z$ being the common normal of the contact plane. Let the gap between the two bodies be an axisymmetric convex function $f(r = \sqrt{x^2+y^2})$. If the two bodies are elastically similar (or if the contact is completely frictionless), i.e., if 
\begin{align}
\frac{1-2\nu_1}{G_1} = \frac{1 - 2\nu_2}{G_2},
\end{align}
the normal contact problem between the two bodies can be reduced two the indentation of an elastic half-space with the effective moduli
\begin{align}
\tilde{E} \coloneqq \left(\frac{1 - \nu_1}{2G_1} + \frac{1 - \nu_2}{2G_2}\right)^{-1} \quad , \quad \tilde{G} \coloneqq \left(\frac{2 - \nu_1}{4G_1} + \frac{2 - \nu_2}{4G_2}\right)^{-1}.
\end{align}
The $z$-axis shall point into the elastic half-space, i.e., $z \geq 0$. Then, a pressure distribution 
\begin{align}
p^{\text{H}}(r;a) \coloneqq -\sigma_{zz}^{\text{H}}(r,z=0;a) = \frac{2\tilde{E}}{\pi R}\sqrt{a^2 - r^2}, \quad r \le a,
\end{align}
in the half-space generates a stress field $\sigma_{ij}^{\text{H}}(x,y,z;a)$, which has been given in explicit form by Huber \cite{Huber1904} and Hamilton \cite{Hamilton1983}, and which will therefore not be re-iterated here for reasons of space. Accordingly, due to linearity, a pressure distribution
\begin{align}
p^{\text{FP}}(r;a) \coloneqq -\sigma_{zz}^{\text{FP}}(r,z=0;a) \coloneqq d^{\text{FP}} R \frac{\partial p^{\text{H}}(r;a)}{\partial (a^2)} = \frac{d^{\text{FP}}\tilde{E}}{\pi}\frac{1}{\sqrt{a^2 - r^2}}, \quad r < a, \label{Eq_Pressure_FlatPunch}
\end{align}
generates a stress field
\begin{align}
\sigma_{ij}^{\text{FP}}(x,y,z;a) = d^{\text{FP}} R \frac{\partial \sigma_{ij}^{\text{H}}(x,y,z;a)}{\partial (a^2)}.
\end{align}
However, as is widely known, the pressure distribution \eqref{Eq_Pressure_FlatPunch} corresponds to the indentation of the elastic half-space by a rigid cylindrical flat punch with the indentation depth $d^{\text{FP}}$ \cite{Boussinesq1885}.

Now consider the indentation by an arbitrary axisymmetric indenter with the profile $f(r)$, starting with a contact radius $\tilde{a} = 0$ and ending with the contact radius $\tilde{a} = a$. The corresponding evolution of the indentation depth is given by
\begin{align}
\tilde{d} = g(\tilde{a}),
\end{align}
with the equivalent plain profile (\cite{Popovetal2019}, section 11.2)
\begin{align}
g(\tilde{a}) = \tilde{a} \int \limits_0^{\tilde{a}} \frac{f'(r)}{\sqrt{\tilde{a}^2-r^2}}\text{d}r,
\end{align}
where the prime denotes the first derivative with respect to the given argument. As has been shown, this indentation procedure can be understood as a series of incremental flat punch indentations by the indentation depths $\text{d}g(\tilde{a})$, with increasing radii $\tilde{a}$. Hence, the complete stress state after the indentation procedure is given by
\begin{align}
\sigma_{ij}^{\text{AS}}(x,y,z;a) = R \int \frac{\partial \sigma_{ij}^{\text{H}}(x,y,z;a)}{\partial (a^2)}\text{d}g(\tilde{a}) =  R \int \limits_0^a \frac{\partial \sigma_{ij}^{\text{H}}(x,y,z;\tilde{a})}{\partial (\tilde{a}^2)}g'(\tilde{a})\text{d}\tilde{a}.
\end{align}
This result has in slightly different form already been reported by Forsbach \cite{Forsbach2020}. To get rid of the derivative of the known solution for the parabolic contact, we introduce the substitutions
\begin{align}
\xi \coloneqq \tilde{a}^2, \quad \alpha \coloneqq a^2, \quad \hat{g}(\xi) \coloneqq g(\tilde{a}), \quad \hat{\sigma}_{ij}(x,y,z;\xi) \coloneqq \sigma_{ij}(x,y,z;\tilde{a}).
\end{align}
As
\begin{align}
g'(\tilde{a})\text{d}\tilde{a} = \text{d}g(\tilde{a}) = \text{d}\hat{g}(\xi) = \hat{g}'(\xi)\text{d}\xi,
\end{align}
integrating by parts then results in
\begin{align}
\hat{\sigma}_{ij}^{\text{AS}}(x,y,z;\xi) = R\left[\hat{\sigma}_{ij}^{\text{H}}(x,y,z;\alpha)\hat{g}'(\alpha) - \int \limits_0^{\alpha} \hat{\sigma}_{ij}^{\text{H}}(x,y,z;\xi) \hat{g}''(\xi)\text{d}\xi \right],
\end{align}
i.e., the stress state for the general axisymmetric Boussinesq problem has been reduced to an appropriate superposition of states for the respective parabolic contact problem.

\subsection{Tangential contact}

\begin{figure}[ht!]
\centering
\includegraphics[width = .8\textwidth]{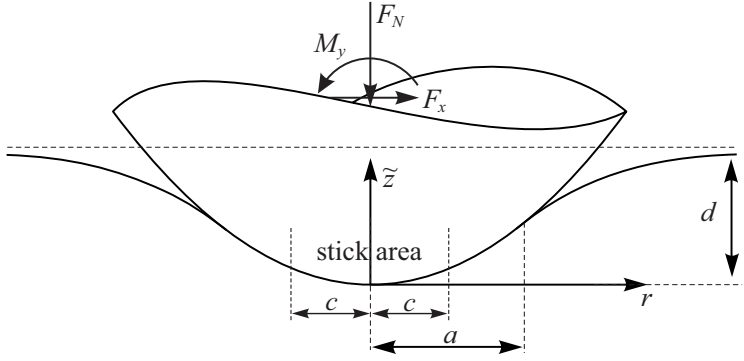}
\caption{Scheme of the axisymmetric tangential contact problem}
\label{Figure1}
\end{figure}

The derivation for the stress state resulting from the tangential tractions can be done in completely similar fashion. Without loss of generality the direction of the tangential loading shall be $x$. In addition to the assumptions made in the previous subsection, we postulate the validity of a local Amontons-Coulomb friction law with a constant coefficient offriction $\mu$ and neglect the lateral surface displacements resulting from the tangential tractions. Under this approximation, the tangential tractions are uni-directional and axisymmetrically distributed. The contact area generally consists of an inner stick area with the radius $c$ and a slip annulus $c < r \le a$. Note that the error of this approximation is usually small as was demonstrated by Munisamy \textit{et al.} \cite{Munisamyetal1994} via a self-consistent numerical solution for the Hertzian contact problem under shear loading. A sketch of the tangential contact problem is shown in figure \ref{Figure1} (note that the surface normal axis pointing \textit{outwards} is referenced as $\tilde{z}$).

Within the framework of the stated assumptions, a shear stress distribution 
\begin{align}
\tau^{\text{H}}(r;a) \coloneqq -\sigma_{xz}^{\text{H},\tau}(r,z=0;a) = \frac{2 \mu \tilde{E}}{\pi R}\sqrt{a^2 - r^2}, \quad r \le a,
\end{align}
generates a stress field $\sigma_{ij}^{\text{H},\tau}(x,y,z;a)$, which has been given in explicit form by Hamilton \cite{Hamilton1983}\footnote{Note that there are some small typing errors in Hamilton's paper; in the solution for $\sigma_{xx}$ it should be $x^2z^2/S$ instead of $x^2z^2/3$. Also, the stress $\sigma_{xy}$ inside the contact area for the normal load is wrong, but can be reconstructed easily considering the system's symmetry.}. Accordingly, due to linearity, a pressure distribution
\begin{align}
\tau^{\text{FP}}(r;a) \coloneqq u_x^{(0)} R \frac{\tilde{G}}{\mu \tilde{E}} \frac{\partial \tau^{\text{H}}(r;a)}{\partial (a^2)} = \frac{u_x^{(0)}\tilde{G}}{\pi}\frac{1}{\sqrt{a^2 - r^2}}, \quad r < a, \label{Eq_Shear_FlatPunch}
\end{align}
generates a stress field
\begin{align}
\sigma_{ij}^{\text{FP},\tau}(x,y,z;a) = u_x^{(0)} R \frac{\tilde{G}}{\mu \tilde{E}} \frac{\partial \sigma_{ij}^{\text{H},\tau}(x,y,z;a)}{\partial (a^2)}.
\end{align}
However, the shear stress distribution \eqref{Eq_Shear_FlatPunch} corresponds to the tangential rigid body displacement of the contact circle $r \le a$ by $u_x^{(0)} = h(a)$ (\cite{Johnson1985}, section 3.7). Therefore, in complete analogy to the normal problem detailed above, the stress state for the axisymmetric indenter, resulting from the tangential contact tractions, is given by
\begin{align}
\hat{\sigma}_{ij}^{\text{AS},\tau}(x,y,z;\xi) = \frac{\tilde{G} R}{\mu \tilde{E}}\left[\hat{\sigma}_{ij}^{\text{H},\tau}(x,y,z;\alpha)\hat{h}'(\alpha) - \int \limits_0^{\alpha} \hat{\sigma}_{ij}^{\text{H},\tau}(x,y,z;\xi) \hat{h}''(\xi)\text{d}\xi \right], \label{Eq_Tang_StressSup}
\end{align} 
with the superposition function $h(\tilde{a}) = \hat{h}(\xi)$, which yet has to be determined.

For the Cattaneo-Mindlin problem (constant normal force and subsequently applied increasing tangential force), the determination of $h$ has been demonstrated by Popov \textit{et al.} (\cite{Popovetal2019}, section 11.5). Consider the following indentation history: first the indenter is just pressed into the elastic medium (without tangential displacement) until a contact radius $\tilde{a} = c$ is reached; after that, every further indentation is associated with an incremental tangential rigid body displacement of the current contact area by
\begin{align}
\text{d}h(\tilde{a}) = \frac{\mu \tilde{E}}{\tilde{G}}\text{d}g(\tilde{a}).
\end{align}
It has been shown that this procedure of generating the contat configuration will fulfill all boundary conditions of the axisymmetric Cattaneo-Mindlin problem. Hence, in this case, the superposition function reads
\begin{align}
h^*(\tilde{a};a,c) = \frac{\mu \tilde{E}}{\tilde{G}}\left[g(\tilde{a}) - g(c)\right], \quad c < \tilde{a} \le a,
\end{align}
where the star shall imply, that this is a basis solution for the elementary Cattaneo-Mindlin loading history.

For general loading histories, one can either refer to Jäger's solution \cite{Jaeger1993} in terms of a superposition of basis solutions -- due to linearity, the superposition will, of course, also be correct for the function $h(\tilde{a})$ -- or one uses the one-dimensional tangential displacements $u_x^{1\text{D}}$ in a model of the system within the framework of the method of dimensionality reduction \cite{PopovHess2014}, which are related to the function $h$ via
\begin{align}
u_x^{1\text{D}}(\tilde{a}) = h(a) - h(\tilde{a}) \quad , \quad h(\tilde{a}) = u_x^{1\text{D}}(0) - u_x^{1\text{D}}(\tilde{a}).
\end{align}
Both solution procedures for arbitrary loading histories are equivalent and can be transformed into one another \cite{Willert2021}. Note that the same holds true for the analysis of general 2D loading of axisymmetric contacts based on the method of memory diagrams \cite{Aleshinetal2015}.

\section{Numerical example}\label{sec3}

\begin{figure}[ht!]
\centering
\includegraphics[width = .7\textwidth]{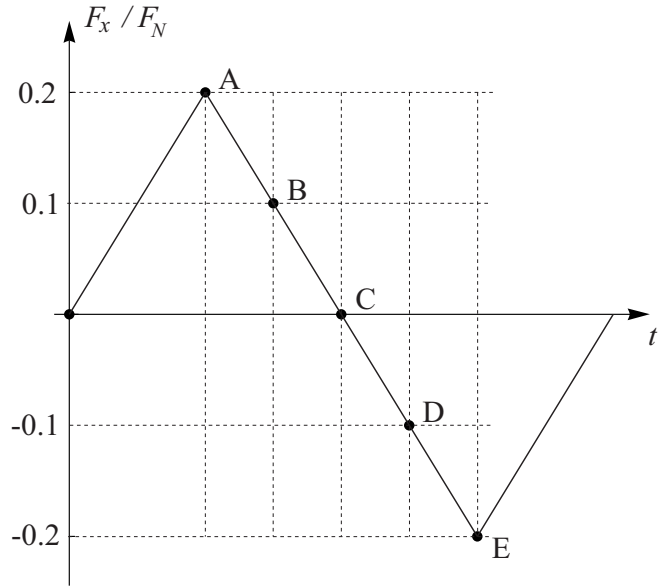}
\caption{Scheme of the loading history for the exemplary calculation.}
\label{Figure2}
\end{figure}

As an example, let us consider the indenter profile $f(r) = r^4/V$, with some constant $V > 0$ with the dimension of a volume. The equivalent profile is $g(\tilde{a}) = (8\tilde{a}^4)/(3V)$, the Poisson ratio shall be $\nu = 0.3$, the friction coefficient $\mu = 0.5$ and the normal load shall be fixed such that $a = V^{1/3}/8$, which corresponds to an average contact pressure of $p_0 = \tilde{E}/(120 \pi)$. The tangential force shall oscillate in the form depicted in figure \ref{Figure2}. We would like to know the distribution of the von-Mises equivalent stress and the highest principal stress in the $xz$-plane for the points A to E on the first  unloading and reversed loading segments of the loading history.

First, the pure contact problem needs to be solved. From Jäger's general oblique loading solution for parabolic contact \cite{Jaeger1993} and the principle by Ciavarella \cite{Ciavarella1998} and Jäger \cite{Jaeger1998} it is clear that, within the framework of the Hertz-Mindlin approximation, the tangential contact stress distribution during the A-E segment of the loading history is given by
\begin{align}
\tau(r) = \tau^*(r;a,c_{\text{min}}) - 2\tau^*(r;a,c), \label{Eq_Jaeger_Sup}
\end{align}
where the star once again denotes the basic solution for the given indenter profile and the elementary Cattaneo-Mindlin loading: $c_{\text{min}}$ is the minimum stick radius at the reversal point A and $c$ is the "current" stick radius. Due to linearity, the same superposition will be correct for the characteristic function $h$. Hence,
\begin{align}
h(\tilde{a}) = h^*(\tilde{a};a,c_{\text{min}}) - 2h^*(\tilde{a};a,c). \label{Eq_Jaeger_Sup_h}
\end{align}
Yet, the stick radii are unknown. The contact solution for the elementary loading yields (\cite{Popovetal2019}, subsection 4.6.4)
\begin{align}
F_x^*(a,c) = \mu F_N\left[1 - \left(\frac{c}{a}\right)^5\right],
\end{align}
and therefore, at the reversal point A,
\begin{align}
c_A = c_{\text{min}} = a\left(1 - \frac{0.2}{\mu}\right)^{1/5}.
\end{align}
Moreover, the superposition \eqref{Eq_Jaeger_Sup} is obviously also correct for the respective tangential forces. Hence, for the loading segment A-E,
\begin{align}
F_x = F_x^*(a,c_{\text{min}}) - 2F_x^*(a,c),
\end{align}
which provides the following stick radii at the points B to D of the loading curve,
\begin{align}
c_B = a\left(1 - \frac{0.1}{2\mu}\right)^{1/5} \quad , \quad c_C = a\left(1 - \frac{0.1}{\mu}\right)^{1/5} \quad , \quad c_D = a\left(1 - \frac{0.3}{2\mu}\right)^{1/5},
\end{align}
and, trivially, $c_E = c_A$. After that, \eqref{Eq_Jaeger_Sup_h} provides the relevant function $h$ at the different instances of loading, which via the general stress superposition rule \eqref{Eq_Tang_StressSup} completely determines the full stress state beneath the contact.

\begin{figure}[ht!]
\centering
\includegraphics[width = .82\textwidth]{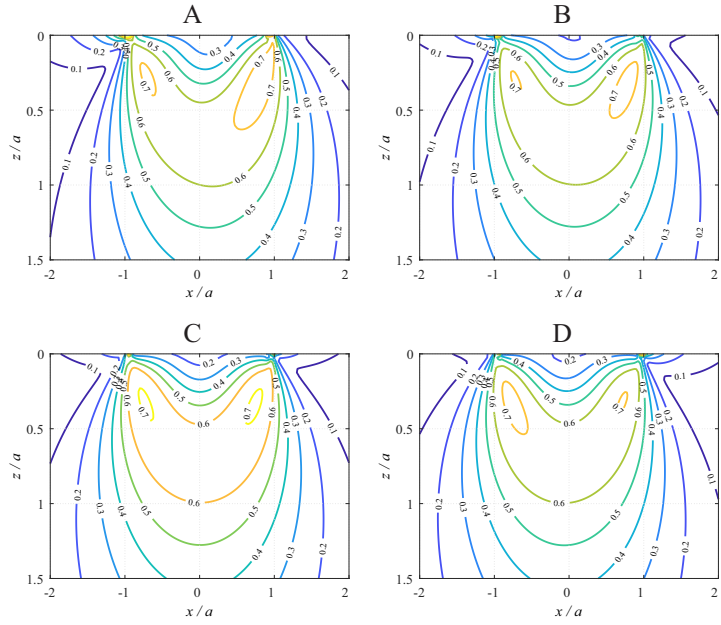}
\caption{distribution of the von-Mises equivalent stress in the $xz$ plane, normalized for the average contact pressure, at the loading points A to D for Poisson ratio $\nu = 0.3$ and friction coefficient $\mu = 0.5$}
\label{Figure3}
\end{figure}

In figure \ref{Figure3} the calculated distributions in the $xz$ plane for the von-Mises equivalent stress are shown (note that the calculation is practically instant, as only trivial one-dimensional integrals have to be solved). The states B and D are mirrored to each other with respect to the $z$ axis; the same will, of course, also be true for the states A and E (the solution at the second reversal point E is therefore ommited here for reasons of space).

\begin{figure}[ht!]
\centering
\includegraphics[width = .82\textwidth]{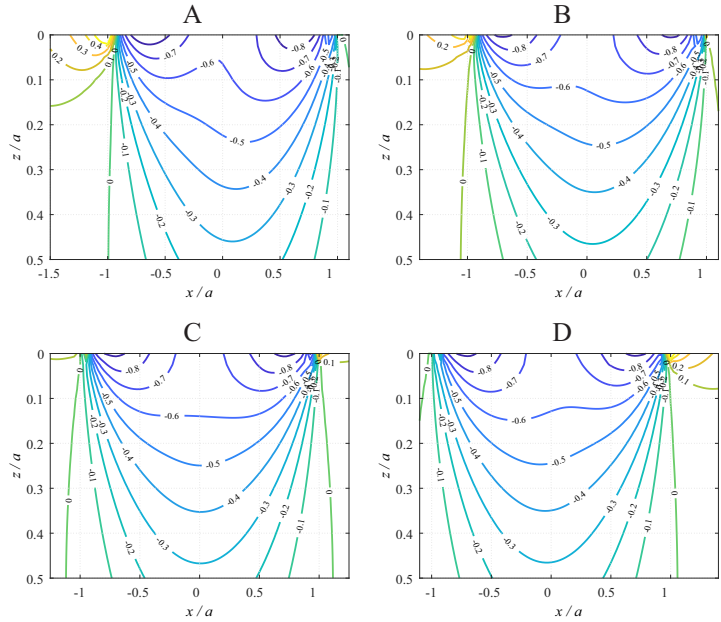}
\caption{distribution of the highest principal stress in the $xz$ plane, normalized for the average contact pressure, at the loading points A to D for Poisson ratio $\nu = 0.3$ and friction coefficient $\mu = 0.5$}
\label{Figure4}
\end{figure}

In figure \ref{Figure4} the respective solutions for the largest principal stress are shown. As expected, at the point A, the largest principal stress is positive at what up until reversing motion has been the trailing contact edge. That is also reversed during the unloading and reversed loading stage.

\section{Conclusions}\label{sec4}

Improving a recent idea for the determination of the full stress state beneath axisymmetric elastic contacts under normal and uni-directional tangential loading via the superposition of respective flat-punch solutions, the determination via the superposition of solutions for parabolic contact has been shown. The numerical implementation of the latter one is even more straightforward, because the lengthy derivatives of the parabolic solution, which have been used in the first approach, can be avoided. Moreover, the implementation, at least if the derivative $\hat{g}''(\xi)$ is not singular at $\xi = 0$, is significantly more stable due to the absence of stress singularities for the parabolic contact. To the author's best knowledge, the presented method provides the fastest and simplest way to determine the stress state under axisymmetric tangential contacts within the framework of the Hertz-Mindlin approximation. It can, e.g., be used to comprehensively analyse the influence of profile geometry on fretting fatigue in dry, oscillating axisymmetric contacts.

\section{Acknowledgements}

This work was supported by the German Research Foundation under the project number PO 810/66-1.

\printbibliography

\end{document}